\documentstyle[preprint,aps,epsf,eqsecnum,floats]{revtex}
\tighten
\begin{document}

{\tighten
\preprint{\vbox{\hbox{JHU--TIPAC--950004}
\hbox{hep-ph/9502248}}}

\title{Instantons and the endpoint of the lepton energy spectrum
in charmless
semileptonic $B$ decays}
\author{Adam F.~Falk and Alexander~Kyatkin}
\address{Department of Physics and Astronomy,
The Johns Hopkins University\\
3400 North Charles Street,
Baltimore, Maryland 21218 U.S.A.\\
{\tt falk@planck.pha.jhu.edu}\\
{\tt kyatkin@fermi.pha.jhu.edu}}

\date{June 1995}

\maketitle
\begin{abstract}
A recent calculation by Chay and Rey has shown that instantons may
make a
significant contribution to the lepton energy spectrum near its
endpoint.
Using an ansatz borrowed from the study of high energy baryon
number violating
processes, we investigate whether these corrections could spoil
the relation
between the nonperturbative contributions to this spectrum and to
the photon
energy spectrum in radiative $B$ decays.  We find, in general,
that this
universality may well fail unless the spectrum is smeared over a
region which
is considerably larger than had previously been thought necessary.
This result
affects the possibility of performing a reliable measurement of
$V_{ub}$ using
inclusive decays.
\end{abstract}

}

\section{Introduction}

There has been considerable recent interest in the study of the
endpoint
spectra of charmless semileptonic and radiative $B$ meson decays.
This effort
is motivated by the fact that in order to remove overwhelming
backgrounds due
to decays to charmed states, stringent cuts must be applied to the
data, which
restrict the experimental analysis to within a few hundred MeV of
the kinematic
endpoint.  Hence it is crucial to understand in as much detail as
possible the
theoretical shape of the lepton or photon energy spectrum in the
endpoint
region, if one is to use these processes to extract reliably
short-distance
couplings such as Kobayashi-Maskawa matrix elements.

The current theoretical analysis makes use of the Operator Product
Expansion
(OPE) and the Heavy Quark Effective Theory
(HQET)~\cite{CGG,Shifetal,others}.
Within this context one may compute a variety of corrections to
the simple free
quark decay picture, both perturbative ${\cal O}(\alpha_s^n)$ and
nonperturbative ${\cal O}(\Lambda^n/m_b^n)$ in origin.  An
important result of
this analysis is that the leading nonperturbative power
corrections in the
endpoint region may be resummed into a universal shape function,
which
describes the distribution of the light-cone momentum of the $b$
quark inside
the $B$ meson~\cite{Neubert,Shif2}.  Since the same
nonperturbative matrix
elements describe the endpoints of both the lepton spectrum in
charmless
semileptonic decays and the photon spectrum in radiative decays,
it is possible
in principle to use a measurement of the photon spectrum to
predict the shape
of the lepton spectrum and thereby allow for a model-independent
measurement of
$V_{ub}$~\cite{Neubert,FJMW}.

This relation is useful, of course, only if the dominant
contribution to the
shape of the lepton endpoint spectrum actually comes from the
nonperturbative
power corrections.  One possible source of trouble is radiative
corrections,
which near the endpoint suppress the theoretical cross-section by
a factor
$\exp[-{2\alpha_s\over3\pi}\ln^2(1-y)]$, where $y=2E_\ell/m_b$ is
the scaled
lepton energy.  For $y$ sufficiently close to 1, this Sudakov
suppression
dominates the theoretical spectrum; whether this is true over the
entire
experimentally allowed window is less clear~\cite{FJMW}.  There
has been recent
progress toward resumming the leading and subleading Sudakov
logarithms, which
would reduce considerably the uncertainty due to this
effect~\cite{KS}.

Another potential source of large corrections near $y=1$ is
instanton effects.
Chay and Rey~\cite{CR} have recently computed the one-instanton
contribution to
inclusive $B$ decays, in the dilute gas approximation (DGA).
Their conclusion
was that for charmless semileptonic decays this contribution
diverges severely
at $y=1$, while it is small and under control for radiative
decays.
Unfortunately, their suggestion that one regulate this divergence
by
considering the energy spectrum only in the region $y<1-\delta$,
where
$\delta\approx0.16\sim0.20$, is not necessarily practical, given
that the
experimental analysis is restricted kinematically to the region
$y\agt0.85$.
In the region of experimental interest, the effect of instantons
is potentially
large and dangerous.  Unfortunately, it is also the region in
which the DGA
begins to break down and multi-instanton processes become
important.

In this paper we will investigate whether instantons spoil the
relationship
between the radiative and semileptonic endpoint spectra in a way
that
necessarily destroys its phenomenological usefulness.  We will
adopt an
approach used in similar situations in the study of baryon number
violation in
high energy collisions~\cite{EKRSW}, in which we use the
one-instanton result
as a guide to an ansatz for the multi-instanton contribution.
This ansatz
contains a small number of physical parameters, and we will
investigate the
size of instanton effects as a function of these parameters.  We
will consider
both the overall magnitude of the instanton contribution and the
order-by-order
behaviour of its moments, as compared to the nonperturbative
corrections which
arrive from higher order terms in the OPE.

The limitations of such an approach are clear.  We will be dealing
not with the
true multi-instanton cross-section, which has not been computed,
but with an
ansatz which has been inspired by a one-instanton calculation
which is valid in
a different region.  Nonetheless, we will come to conclusions
which we believe
are robust, and which indicate that large instanton corrections to
the shape of
the endpoint spectrum may be difficult to avoid.

\section{The one-instanton calculation}

We begin by summarizing the calculation of Chay and Rey~\cite{CR}
of the effect
of a single instanton on the lepton and photon energy spectrum.
In the context
of the OPE, the decay width is determined by the correlator of two
quark
bilinears.  For example, for the process $B\to X_u\ell\nu$, the
differential
decay rate is given by
\begin{equation}
   {\rm
d}\Gamma={G_F^2\over4m_b}|V_{ub}|^2W^{\mu\nu}L_{\mu\nu}{\rm
d(P.S.)}\,,
\end{equation}
where $L_{\mu\nu}{\rm d(P.S.)}$ is the product of the lepton
matrix elements
with a lepton phase space measure, and
\begin{equation}\label{Wmunu}
   W^{\mu\nu}=-2\,{\rm Im}\left\{ i\int{\rm d}^4x e^{iq\cdot x}
   \langle B\,|\, T\{\bar b\gamma^\mu(1-\gamma^5)u(x),
   \bar u\gamma^\nu(1-\gamma^5)b(0)\}\,|\,B\rangle\right\}
\end{equation}
describes the interactions of the quarks~\cite{CGG}.  The
correlator is
developed in a simultaneous expansion in $\alpha_s$ and the off
shell momentum
of the $u$ quark, which is of order $m_b$ everywhere but at the
boundaries of
phase space.  In terms of the scaled variables $y=2p_b\cdot
k_\ell/m_b^2$
($=2E_\ell/m_b$ in the $B$ rest frame) and $\hat
s=(k_\ell+k_\nu)^2/m_b^2$,
these boundaries are at $y=\hat s$ and $y=1$.

The calculation of the correlator~(\ref{Wmunu}) in the dilute
instanton
background gives the instanton contribution to the decay
width~\cite{CR}.  The
instanton contribution enters as a contribution to the
coefficients of the
operators which appear in the OPE.  The computation involves an
integration
over the instanton size $\rho$, which diverges in the infrared.
Chay and Rey
deal with this divergence by expanding the integrand in $1/\rho$
and
interpreting the finite number of divergent terms as contributions
to the
matrix elements of operators in the OPE.  This is appropriate
insofar as the
divergent contribution of large instantons is presumably regulated
physically
by the infrared growth of the QCD self-coupling.  The terms which
are infrared
convergent and hence calculable are interpreted as contributions
to the
coefficient functions in the OPE.

With some mild additional approximations, Chay and Rey derive an
expression for
the leading one-instanton contribution to the doubly differential
decay width,
\begin{equation}\label{inst1}
   {1\over\Gamma_0}{{\rm d}^2\Gamma_{\rm inst}\over{\rm d}\hat
s{\rm d}y}
   =A\,y^5\,{5\hat s-(1-y)(y-\hat s)\over(1-y)^6(y-\hat s)^5}\,,
\end{equation}
where $\Gamma_0=G_F^2|V_{ub}|^2m_b^5/192\pi^3$.  The constant $A$
depends on
the quark masses, the QCD scale $\Lambda$ and the number of light
flavors, and
is estimated by Chay and Rey to be
\begin{equation}
   A = \left({19.2\,{\rm GeV}\over
m_b}\right)^3\left({\Lambda\over
   m_b}\right)^9\approx6.7\times10^{-8}\,,
\end{equation}
for $\Lambda=350\,$MeV and $m_b=4.5\,$GeV.  Note that the
expression~(\ref{inst1}) scales na\"\i vely as $m_b^{-12}$ and has
strong
divergences as $y\to1$ and $y\to\hat s$.  For the radiative decay
$B\to
X_s\gamma$, under the same approximations, Chay and Rey find the
one-instanton
contribution to the photon energy spectrum,
\begin{equation}\label{inst2}
   {1\over\Gamma_{0,\gamma}}{{\rm d}\Gamma_{{\rm
inst},\gamma}\over{\rm
   d}y_\gamma}
   =\left({26.1\,{\rm GeV}\over m_b}\right)^3\left({\Lambda\over
m_b}\right)^9
   {y_\gamma^3\over(1-y_\gamma)^7}\,,
\end{equation}
where $y_\gamma=2p_b\cdot k_\gamma/m_b^2$ is the scaled photon
energy in the
$B$ rest frame, and $\Gamma_{0,\gamma}$ is the lowest order free
quark
radiative decay width.  Again, the instanton contribution diverges
strongly as
$y_\gamma\to1$.  However, only integrals of Eq.~(\ref{inst2}) are
actually
meaningful, as we discuss below.

The divergent behaviour of the instanton contribution at the edges
of phase
space has a straightforward origin.  Once the infrared divergences
have been
subtracted, the contribution of instantons to the coefficient
functions comes
from small instantons of typical size $\rho\le1/|Q|$, where the
scale $Q$ is
determined by the momentum of the final quark propagating in the
instanton
background.  Since the instanton contribution contains the
suppression factor
$\exp(-2\pi/\alpha_s(Q^2))$, it is important only when
$Q^2/\Lambda^2$ is of
order one, that is, when the invariant mass of the hadronic final
state is of
order the QCD scale rather than the bottom mass.  This occurs at
the boundaries
of phase space, where the final state light quark is driven to its
mass shell.
Hence this is the region where instanton effects become not only
significant,
but divergent.

An alternative approach to this calculation would be to cut off
all instantons
with $\rho\agt1/m_b$, which would suppress the one-instanton
contribution by
$(\Lambda/m_b)^9$, independent of $y$.  Such a cutoff might be
natural for a
diagram in which all propagators are in the instanton background,
such as the
calculation of the polarization operator in $e^+e^-$ annihilation.
However,
for semileptonic and rare $B$ decays the instanton insertions are
only in one
propagator, and the choice of cutoff should be governed solely by
the
kinematics of this light quark, not by the total energy released
in the decay.
Effectively, we cut off the integral over $\rho$ at
$\rho\sim1/|Q|$, leading to
a differential rate which is unsuppressed at the lepton energy
endpoint.

Essentially, the question is whether the integral over $\rho$
should be
performed before or after the integration over loop momenta.  If
one performs
the loop integration first, then the only remaining {\it
external\/} momentum
scale is set by $m_b$, and a cut on $\rho\agt1/m_b$ would seem
natural.  If,
however, we perform the integral over $\rho$ first, then a loop
momentum-dependent cutoff prescription such as we use becomes
possible, and in
this case, for the reasons discussed above, is preferred.

In the case of $B\to X_s\gamma$ decays, it is possible to regulate
the
divergence in Eq.~(\ref{inst2}) in such a way that the
contribution of
instantons to the {\it total} decay rate is finite and, in fact,
negligibly
small~\cite{CR}.  The situation is similar to that of the
contributions of
Sudakov double logarithms to this process~\cite{Shif2}; so long as
the cut
$y>y_c$ on the photon momentum is not too stringent ($y_c\alt0.85$
will do), it
is possible to analytically continue the phase space integral away
from any
resonance region.  Once this has been done, the one-instanton
contribution is
small, and calculable, everywhere.  Hence, the one-instanton
result gives a
reliable estimate of instanton contributions to integrated
quantities such as
the total decay width.

Unfortunately, this procedure will not work for $B\to X_u\ell\nu$
decays.  In
this case one must perform an integration in the variable
$p_b\cdot
(k_\ell+k_\nu)/m_b^2$, and near the boundary of phase space the
contour cannot
be deformed away from the resonance region~\cite{CGG,CR}.  (This
happens both
because the endpoint of the integration is a function of $y$ and
$\hat s$ and
cannot be adjusted by hand, and because in certain regions of
phase space the
contour is pinched between two cuts.)  As a result, in this regime
we do expect
a large instanton contribution.

Of course, in this region the OPE itself breaks down, as
corrections from
operators of higher twist become important.  However, the
one-instanton
correction has such severe divergences in this region that
instantons become
important even when $Q^2>\Lambda^2$, so it is not unreasonable to
expect that
the OPE analysis gives the correct order of magnitude of instanton
effects.

Even in the absence of nonuniversal higher twist contributions,
instantons
contribute differently to semileptonic and rare $B$ decays.  The
reason is
essentially the dependence, in the semileptonic case, of the
boundary point of
the contour of integration on the kinematics of the leptons.  Such
a dependence
does not affect universality in the absence of instanton effects,
because the
OPE is an expansion in inverse powers of $Q^2$ and therefore has a
pole
behaviour, which is insensitve to the shape of the contour of
integration.
Instanton zero modes, by contrast, give  $\log Q^2$ contributions
to
coefficient functions, so instanton corrections depend explicitly
on the
position of the boundary point of the contour.

\section{The multi-instanton ansatz}

The calculation discussed in the previous section shows that the
one-instanton
contribution to the semileptonic differential decay rate becomes
of the same
size as the lowest order result in the region $y\agt0.9$, and
diverges in the
limit $y\to1$.  In this region, then, we cannot trust any more the
one-instanton result, and we must include multi-instanton
corrections as well.
Unfortunately, there is not at present a technology for performing
such
calculations.

The situation is similar to one obtaining in the study of high
energy
collisions, where the one-instanton correction to the total cross
section grows
exponentially with energy and violates unitarity in the multi-TeV
region.
Possible ways of treating this problem have been discussed widely
in the
literature (for a review, see Ref.~\cite{Mattis}).  Most likely,
multi-instanton corrections stop this dangerous growth and
unitarize the
amplitude at high energies.  While the detailed behaviour of the
multi-instanton contribution is, of course, not known, one makes a
hypothesis
as to its qualitative form.  It is assumed that the
instanton-mediated cross
section has a threshold behaviour:  it is dominated by the
one-instanton
contribution below the threshold, and hence exponentially
suppressed; it
reaches the unitarity bound in the threshold region; and it stays
almost
constant above threshold, in the multi-instanton regime.  These
properties lead
to an ansatz for the full instanton contribution, which is a step
function with
support above the threshold~\cite{EKRSW}.  This ansatz has two
parameters: the
width of the step, corresponding to the position of the threshold,
and the
height, corresponding to the strength of the unitarized amplitude.
While such
an ansatz is obviously extremely crude, it incorporates the one
useful piece of
information which may be extracted from the one-instanton
calculation:  the
energy at which the instanton contributions become large.  Because
of the rapid
rise of the one-instanton contribution with energy, this threshold
is actually
predicted fairly reliably.

We will follow an analogous procedure in our discussion of the
instanton
correction to the semileptonic spectrum.  We will arrive at an
ansatz for the
multi-instanton contribution which is equally crude, but which we
hope will
again incorporate correctly the information provided by the
one-instanton
calculation.  To try to account as honestly as possible for the
large and
uncontrolled uncertainty in our ansatz, we will vary the
parameters which
define it over ranges which, in our opinion, are quite generous.

We note briefly the claim in the literature~\cite{ZMS} that the
one-instanton
result actually unitarizes ``prematurely'', with its growth
stabilized when it
is still exponentially small.  This proposal is still somewhat
controversial~\cite{Mattis,AK}; if correct, it will result in a
strong
suppression of all instanton effects.  We will not address this
issue further,
except to note that in our ansatz we allow for a significant
variation in the
overall normalization.  This normalization can be taken in
principle to include
the effect of premature unitarization, although if the phenomenon
is real then
it may lead to a stronger suppression than we consider below.

\begin{figure}
\epsfxsize=8cm
\hfil\epsfbox{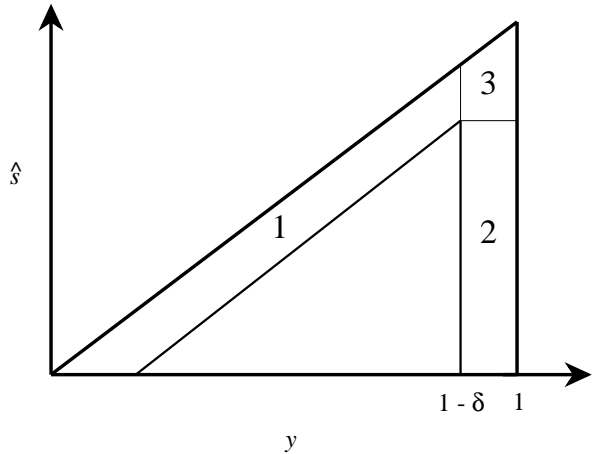}\hfill
\caption{The one-instanton calculation is reliable only within the
inner
triangle.  In regions 1, 2 and 3, we replace the one-instanton
result with the
multi-instanton ansatz, as explained in the text.}
\label{phasespace}
\end{figure}

We begin with the expression (\ref{inst1}) for the one-instanton
contribution
to the doubly differential decay rate.  This expression is
severely divergent
as $\hat s\to y$ and $y\to1$, and particularly so when both limits
are taken
simultaneously.  Let us suppose, then, that we believe the
one-instanton
calculation only over the region $R=\{0\le\hat s\le y-\delta, \
\delta\le
y\le1-\delta\}$, where $\delta\ll1$.  Outside of $R$,
multi-instanton
configurations probably regularize the otherwise divergent decay
rate, and the
largest instanton contributions to the rate actually come from
this region at
the boundary of phase space.  Hence we replace the one-instanton
result by a
step function ansatz, as shown in Fig.~\ref{phasespace}.
  In region 1 of Fig.~\ref{phasespace}, we take as our ansatz
Eq.~(\ref{inst1}), with $\hat s=y-\delta$; in region 2, we take
Eq.~(\ref{inst1}) with $y=1-\delta$; and in region 3, we take
Eq.~(\ref{inst1})
with $y=1-\delta$ and $\hat s=1-2\delta$.  Elsewhere, the
instanton-mediated
decay rate is dominated by the small one-instanton contribution,
and is taken
to vanish.

Next, we integrate this ansatz over $\hat s$ to get a preliminary
ansatz for
${\rm d}\Gamma_{\rm inst}/{\rm d} y$.  To within a factor of
two\footnote{We
will soon vary the rate over a range much larger than this, so the
approximations involved in obtaining this simple and convenient
expression are
relatively harmless.} for $\delta\alt0.3$, we find the simple
result
\begin{equation}\label{ansatzprelim}
  {1\over\Gamma_0}{{\rm d}\Gamma_{\rm inst}\over{\rm d}y}\approx
  \left\{\begin{array}{ll}
         5Ay^5/(1-y)^6\delta^4 & \mbox{   if $y<1-\delta$} \\
         5A(1-\delta)^5/\delta^{10} & \mbox{   if $1-\delta\le
y<1$}
         \end{array}
  \right.
\end{equation}
Since the function falls steeply for $y<1-\delta$, we make the
further
simplification of setting the function to zero in this region.
The final form
for our ansatz, then, is a step function which takes the form
\begin{equation}\label{ansatz}
{1\over\Gamma_0}{{\rm d}\Gamma_{\rm inst}\over{\rm d}y}=
   {\nu\over\delta}\big\{\Theta(1-y)-\Theta(1-y-\delta)\big\}\,.
\end{equation}
The height $\rho=\nu/\delta$ of the step is given by the second
case of
Eq.~(\ref{ansatzprelim}).

We stress that this ansatz for the multi-instanton contribution to
the width is
extremely crude.  However, it does contain certain useful
information gleaned
from the one-instanton result.  Because of the extremely strong
dependence of
$\rho$ on $\delta$, we actually find a reasonable constraint on
the width.  We
assume that multi-instanton contributions come in and regularize
the width when
$\rho$ gets larger than some value $\rho_0$, and let us take the
very generous
range ${1\over10}<\rho_0<10$.  Then we find, from the expression
(\ref{ansatzprelim}), that $0.16<\delta<0.24$.  This is the range
of $\delta$
which we will allow in Eq.~(\ref{ansatz}).

It is somewhat more difficult to set a reasonable range for the
area $\nu$.  We
would like to allow for the significant uncertainty in the
derivation of
Eq.~(\ref{ansatzprelim}), without losing entirely the strong
dependence of
$\nu=\rho\delta$ on $\delta$, which is physical.  Our prescription
will be to
introduce an {\it ad hoc\/} multiplicative factor in the
normalization of the
ansatz, and to consider a wide variation in its value.  Hence we
will
consider the functions
\begin{equation}\label{cdefs}
  \nu_i=\rho_i\delta=c_i\cdot5A{(1-\delta)^5\over\delta^9}\,,
\end{equation}
where $c_1=1$, $c_2={1\over10}$, and $c_3={1\over100}$.  These
three functions
cover a variation of two orders of magnitude in the true size of
the
multi-instanton contribution, compared with our na\"\i ve ansatz
(\ref{ansatz}).  Combined with the restriction $0.16<\delta<0.24$
obtained
above, we find $0.0002<\nu<1.6$.  While the upper limit is not to
be taken
seriously, we believe that $\nu\agt0.0002$ represents a reasonable
lower limit
on the size of multi-instanton effects.

\section{Implications for universality}

Given this ansatz for the multi-instanton contribution, what are
the
implications for the measurement of $V_{ub}$?  When instanton
effects are
included, do they dominate the shape of the endpoint spectrum, or
is this shape
still determined by the nonperturbative power corrections?  In
order to frame
this question properly, we must consider both the overall size of
the instanton
contribution and its effect on the moments of the spectrum.

The leading power corrections to the lepton energy spectrum may be
expanded in
a series of singular functions near $y=1$,
\begin{equation}\label{exp1}
   {1\over\Gamma_0}{{\rm d}\Gamma_{\rm th}\over{\rm
d}y}=B_0\theta(1-y)
   +B_1\delta(1-y)+B_2\delta'(1-y)+\ldots\,,
\end{equation}
where $B_n\sim(\Lambda/m_b)^n$.  The singular parts of this
expression may be
resummed into a ``shape function'' $S(y)$ of width $\Lambda/m_b$,
so
Eq.~(\ref{exp1}) takes the form~\cite{Neubert}
\begin{equation}\label{shapedef}
   {1\over\Gamma_0}{{\rm d}\Gamma_{\rm th}\over{\rm d}y}=2y\big[
   F(y)\theta(1-y)+F(1)S(y)\big]\,,
\end{equation}
where $F(y)=y(3-2y)+{\cal O}(\alpha_s)$ is a smooth function of
$y$.  It is
convenient to define moments of the shape function,
\begin{equation}\label{momdef}
   M_n^{\rm th}=\int{\rm d}y\,S(y)(y-1)^n\,,
\end{equation}
for $n\ge1$.  The moments scale inversely with the bottom mass,
$M_n^{\rm
th}\sim(\Lambda/m_b)^n$.  If the spectrum is smeared with a
weighting function
of width $\sigma$ near $y=1$, for example
$w(y)\propto\exp[-(y-1)^2/2\sigma^2]$, then the result may be
written as a sum
of these moments,
\begin{equation}\label{momentsum}
   \int{1\over\Gamma_0}{{\rm d}\Gamma_{\rm th}\over{\rm
d}y}\,w(y){\rm d}y=
   \int 2y^2(3-2y)w(y){\rm d}y
   +2\sum_{n=1}^\infty{M_n^{\rm th}\over n!}\,w^{(n)}(1)\,.
\end{equation}
It is this sum which is universal, in the sense that it appears
both in the
expression for the semileptonic endpoint spectrum and that for the
photon
energy spectrum in $B\to X_s\gamma$ transitions.  If the smearing
region
$\sigma$ is of order $\Lambda/m_b$, then all terms in the sum
(\ref{momentsum})
are of the same order in the $1/m_b$
expansion~\cite{Neubert,Shif2,FJMW}.

Instanton corrections appear as an additional term in the shape
function,
\begin{equation}\label{shapeinst}
   S_{\rm inst}(y) ={1\over 2\Gamma_0}\,{{\rm d}\Gamma_{\rm
inst}\over
   {\rm d}y}\,.
\end{equation}
The new contribution to the right hand side of
Eq.~(\ref{momentsum}) takes the
form
\begin{equation}
   \int{1\over\Gamma_0}{{\rm d}\Gamma_{\rm inst}\over{\rm
d}y}\,w(y){\rm d}y
   =2\sum_{n=0}^\infty{M_n^{\rm inst}\over n!}w^{(n)}(1)\,,
\end{equation}
where
\begin{equation}
   M_n^{\rm inst}=\int{1\over2\Gamma_0}{{\rm d}\Gamma_{\rm
inst}\over
   {\rm d}y}\,(y-1)^n{\rm d}y=\int S_{\rm inst}(y)(y-1)^n{\rm
d}y\,,
\end{equation}
in analogy with Eq.~(\ref{momdef}).  Universality will continue to
hold if this
new nonuniversal term is subleading, in some sense, compared to
the universal
series generated by the power corrections.  In fact, there are two
criteria
which we must impose.  The simplest is the condition that the
total instanton
contribution be small compared to the parton model rate.  How
small this ought
to be is somewhat a matter of taste.   As an illustration, let us
take $w(y)$
to be a Gaussian weighting function of width $\sigma$.  Then the
parton model
rate is given by
\begin{equation}
   A_0(\sigma)=\int 2y^2(3-2y)w(y) {\rm d}y\,,
\end{equation}
while the contribution of the instantons depends on the ansatz
parameter
$\delta$,
\begin{equation}
   A_{\rm inst}(\sigma,\delta)=\int{1\over\Gamma_0}{{\rm
d}\Gamma_{\rm inst}
   \over{\rm d}y}\,w(y){\rm d}y \,.
\end{equation}
If we require the inequality
\begin{equation}
   A_{\rm inst}(\sigma,\delta)<\kappa A_0(\sigma)\,,
\end{equation}
for some $\kappa$, then we obtain lower limits on the smearing
region $\sigma$
as a function of $\delta$.  These limits depend also on the
normalization $c_i$
of the multi-instanton ansatz.  In Fig.~\ref{sigmacurves} we show
$\sigma_{min}(\delta)$ for $c_1=1$, $c_2={1\over10}$, and
$c_3={1\over100}$,
and for $\kappa={1\over5}$ (a somewhat strict condition) and
$\kappa=1$ (a much
looser one).

The second criterion is somewhat more subtle, and concerns the
behaviour of the
moments at large values of $n$.  That the moments $M_n^{\rm th}$
scale as
$(\Lambda/m_b)^n$ follows immediately from the form of the
operator product
expansion~\cite{Neubert}.  The instanton contribution, by
contrast, is not an
expansion in the inverse power of a momentum, and there is no
reason for the
instanton moments $M_n^{\rm inst}$ to show such a behaviour.
Indeed, since the
one-instanton result grows so steeply as $y\to1$, its moments at
large $n$ may
be large compared to the parton model result, even if they are
suppressed at
small $n$.  The same is true of our multi-instanton ansatz; a
simple
calculation yields
\begin{equation}\label{instmoms}
   \left|M_n^{\rm
inst}\right|={1\over2(n+1)}\,\delta^n\nu(\delta)\,.
\end{equation}
The instanton moments are only subleading compared to the power
corrections
at large $n$ if $\delta<\Lambda/m_b$.

Since the natural width of the theoretical shape function $S(y)$
is
$\Lambda/m_b$, this is just the condition that the instantons are
concentrated
in a region closer to the endpoint than the smearing given by the
initial $b$
quark motion in the $B$ meson.  If so, then we can neglect not
only the total
instanton contribution but also the contribution of the instantons
to the shape
of the endpoint spectrum.  The estimate of $\delta$ which we
obtained
previously, $0.16<\delta<0.24$, does not always satisfy this
condition, except
for quite large values of the QCD scale, $\Lambda\sim1\,$GeV.

We can resolve this problem only by adjusting the width $\sigma$
of the
smearing function $w(y)$.  Let us suppose that the dependence of
$\sigma$ on
the bottom mass is given by
\begin{equation}\label{epsdef}
   \sigma\sim\left({\Lambda\over m_b}\right)^{1-\epsilon}\,,
\end{equation}
where $\epsilon\ge0$.  As discussed in
Refs.~\cite{Neubert,Shif2,FJMW}, if
$\epsilon=0$ then all of the terms in the series (\ref{momentsum})
are of the
same order in $\Lambda/m_b$, since $M^{\rm
th}_n\sim(\Lambda/m_b)^n$ and
$w^{(n)}\sim1/\sigma^n$.  For $\epsilon>0$, all terms with
$n\agt1/\epsilon$
are suppressed by at least $\Lambda/m_b$ and may be neglected,
since they are
of the same order as other terms which were dropped earlier.
Hence
$\epsilon=0$ provides a lower, but not an upper, limit on the size
of the
smearing region, as a function of $m_b$.

Now suppose that the condition $\delta<\Lambda/m_b$ is not
satisfied.  Then the
higher moments $M_n^{\rm th}$ are power-suppressed compared to
$M_n^{\rm
inst}$, and there is in general an integer $n_{\rm crit}$ such
that for
$n>n_{\rm crit}$ we have $M_n^{\rm inst}>M_n^{\rm th}$.  If we
insist on a
smearing region size corresponding to $\epsilon=0$, for which
moments at all
$n$ contribute equally to the sum (\ref{momentsum}), then this
situation will
obviously lead to trouble.  Since for large $n$ the sum of moments
is dominated
by the instanton terms rather than those from the parton level,
there will be
no useful relation between the smeared spectra in radiative and
rare
semileptonic decays.  Certainly, it will still be possible to use
the observed
photon spectrum in $B\to X_s\gamma$ to predict certain
contributions to the
lepton energy spectrum in $B\to X_u\ell\nu$, but these
contributions will not
be the dominant ones.  Instead, the uncertainty in the shape of
the lepton
energy spectrum will be dominated by the uncertainty in the
contributions of
multi-instanton processes, which, as we have seen, is very large
indeed.

Instead, we must smear over a larger region about $y=1$,
corresponding to an
exponent $\epsilon>0$ in Eq.~(\ref{epsdef}).  If we do so, then
all moments
with $n\agt1/\epsilon$ are subleading and may be ignored.  Hence
if we choose
$\epsilon>\epsilon_{\rm crit}=1/n_{\rm crit}$, then by the time
$M_n^{\rm
inst}=M_n^{\rm th}$, both $M_n^{\rm inst}$ and $M_n^{\rm th}$ may
be neglected.
 In order to suppress the high moments of the instanton
contribution, then, we
are led to require a smearing width $\sigma$ which might be
significantly
larger than previously expected.  The critical smearing depends on
the size of
the multi-instanton ansatz, and is parameterized by a function
$\epsilon_{\rm
crit}(\delta)$.

We can estimate $\epsilon_{\rm crit}(\delta)$ by comparing our
multi-instanton
ansatz to the result which is obtained in the ACCMM
model~\cite{ACCMM}.  The
moments $M_n^{\rm th}$ in this model have been calculated by
Neubert~\cite{Neubert}, with the result
\begin{equation}\label{ACCMMmoms}
  M_n^{\rm th}={n!!\over2^{n+1\over2}}\,\left(p_F\over
m_b\right)^{n+1}\,,
\end{equation}
for $n$ even.  A best fit to the spectrum yields $p_F\approx
230\,$MeV, and we
take $m_b=4.8\,$GeV.  (Because of the symmetries of the model,
$M_n^{\rm th}$
vanishes for $n$ odd.  The model also has the curious feature that
the moments
exhibit an $n!!$ growth, which alters somewhat the condition on
$\delta$.
However, we remind the reader that the model is being used only as
a somewhat
crude comparison to an equally crude ansatz.)  We obtain a value
of $n_{\rm
crit}$ by comparing the moments in Eqs.~(\ref{instmoms}) and
(\ref{ACCMMmoms}),
and from this the critical smearing exponent $\epsilon_{\rm
crit}(\delta)$.
The result in the ACCMM model, however, is that even under the
worst
assumptions, $\epsilon_{\rm crit}\alt{1\over10}$, so this effect
turns out to
be relatively unimportant.  In fact, this is not so surprising:
in our
multi-instanton ansatz, we explicitly have cut off the strong
divergence of the
instanton contribution near the endpoint, so the effect of
instantons on the
shape of the endpoint spectrum is less important than their total
contribution
to the decay rate.

\section{Implications and discussion}

\begin{figure}
\epsfxsize=10cm
\hfil\epsfbox{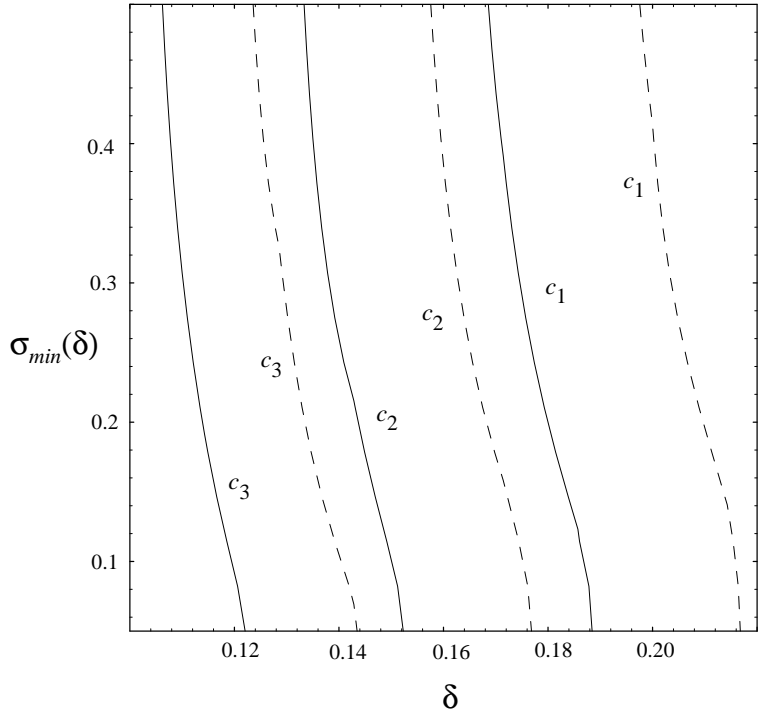}\hfill
\caption{The minimum smearing width $\sigma_{min}(\delta)$ for the
multi-instanton ansatz.  We show curves for the normalizations
$c_1=1$,
$c_2={1\over10}$ and $c_3={1\over100}$ of the multi-instanton
contribution, and
for $\kappa={1\over5}$ (solid curves) and $\kappa=1$ (dashed
curves).}
\label{sigmacurves}
\end{figure}

Do the limits $\sigma_{min}(\delta)$, summarized in
Fig.~\ref{sigmacurves},
indicate that instantons constitute an important effect on the
endpoint
spectrum?  The answer to such a question depends crucially on
precisely how it
is posed.  It is clear that since we are working with an extremely
crude ansatz
for the multi-instanton contribution, no numerical prediction
which results is
to be believed.  However, the real hope of this analysis was to
show that
instanton effects are sufficiently {\it negligible\/} that the
proposal to
measure the shape of the photon energy spectrum in radiative $B$
decays and use
it to predict the shape of the endpoint spectrum in charmless
semileptonic $B$
decays is left unaffected.

As we see from Fig.~\ref{sigmacurves}, this is certainly not the
case.
For example, given the experimental constraints, the smearing
region must
satisfy $\sigma\alt0.2$.  If we now focus on the na\"\i ve
normalization of the
ansatz, $c_1=1$, and on the ``loose'' criterion $\kappa=1$, we see
that only
for $\delta\agt0.18$ is $\sigma_{min}<0.2$.  For $\delta\alt0.18$,
the
instanton effect dominates the nonperturbative shape function over
the
experimental smearing region.  But recall from Section III that
the
one-instanton result indicates that the entire region
$0.16<\delta<0.24$ is
likely to be allowed.  Note that our ansatz hardly has to be
pushed to its
extremes for the relationship between the shapes of the radiative
and charmless
semileptonic spectra to be spoiled.  If we apply the ``strict
criterion''
$\kappa={1\over5}$, then only the curve with the suppression
factor
$c_3={1\over100}$ is acceptable over the entire range
$0.16<\delta<0.24$.  We
conclude, then, that {\it there is no reason to believe that it is
safe to
neglect instantons in the analysis of the lepton energy endpoint
spectrum.}
Hence, we would have no particular confidence in the result of the
proposed
program to measure $V_{ub}$ by comparing endpoint spectra, were it
ever to be
performed.

We stress that our result is interesting and important only in the
{\it
negative\/} sense.  By no means do we claim to have calculated the
effect of
multi-instanton configurations, or even to have estimated them
with any
particular accuracy.  What we have done, instead, is to analyze a
well-motivated multi-instanton ansatz honestly and conservatively.
Within this
ansatz, we have not found instanton effects uniformly to be
negligible, from
which we have concluded that neither are they {\it necessarily\/}
negligible in
the real world.  We note that we would have reached a different
conclusion, had
considerable variation in the ansatz parameters yielded
consistently negligible
results.

Of course, one may take the point of view that instanton effects
could as
easily be tiny as large, and that in the absence of better
evidence one should
proceed on this more hopeful assumption.  However, this na\"\i ve
approach, if
applied to the extraction of $V_{ub}$, would lead to a systematic
uncertainty
which is unknown and probably unknowable.  On the other hand,
perhaps our
ansatz can be refined, or improved, or replaced by something
closer to the
truth.  Eventually, perhaps, effects such as we have considered
here may even
be calculable.  We certainly hope that such advances will one day
prove
instantons to be unimportant to the semileptonic endpoint
spectrum, and the
proposed experimental analysis to be unaffected.  But unless good
reasons arise
for such confidence to be restored, we will consider the
measurement of
$V_{ub}$ by the detailed analysis of the lepton energy endpoint
spectrum to be
intrinsically uncertain and untrustworthy.

\acknowledgements

This work was supported by the National Science Foundation under
Grant
No.~PHY-9404057.  A.F.~also acknowledges the National Science
Foundation for
National Young Investigator Award No.~PHY-9457916, and the
Department of Energy
for Outstanding Junior Investigator Award No.~DE-FG02-094ER40869.

\end{document}